%% file: ms.tex
\documentclass[preprint]{aastex}

\usepackage{longtable}
\usepackage{multirow}
\received{}
\revised{}
\accepted{}
\shorttitle{HIGH VELOCITY COMPACT CLOUDS IN THE SAGITTARIUS C REGION}
\shortauthors{Tanaka et al.}
\tighten

\usepackage{float}
\begin{document}

%%%%%%%%%%%%%%%%%%%%%%%%%%%%%%%%%%%%%%%%%%%%%

\newcommand\pcc{\mathrm{cm^{-3}}}
\newcommand\psc{\mathrm{cm^{-2}}}
\newcommand\kelvin{\mathrm{ K}}
\newcommand\kmps{\ifmmode\mathrm{km\,s^{-1}}\else$\mathrm{km\,s^{-1}}$\fi}
\newcommand\erg{\mathrm{erg}}
\newcommand\ergs{\mathrm{erg}}
\newcommand\pc{\mathrm{pc}}
\newcommand\str{\mathrm{str}}
\newcommand\second{\mathrm{s}}
\newcommand\yr{\mathrm{yr}}
\newcommand\Hz{\mathrm{Hz}}
\newcommand\Msol{{M_\odot}}
\newcommand\Lsol{{L_\odot}}
\newcommand\kB{{k_{\rm B}}}
\newcommand\XHCN{{R_{\rm HCN}}}
\newcommand\XHCOp{{R_{\HCOp{}{}}}}
\newcommand\Tkin{{T_{\rm kin}}}
\newcommand\Ekin{{E_{\rm kin}}}
\newcommand\Etot{{E_{\rm tot}}}
\newcommand\Tex{{T_{\rm ex}}}
\newcommand\Tsys{{T_{\rm sys}}}
\newcommand\Trms{{T_{\rm rms}}}
\newcommand\nH{{n_{\rm H}}}
\newcommand\nHH{{n_{\rm H_2}}}
\newcommand\ncrit{{n_{\rm crit}}}
\newcommand\Column[1]{{N_{#1}}}
\newcommand\dColumn[1]{{\Column{#1}/{{\rm d}v}}}
\newcommand\vlsr{\ifmmode{v_{\rm LSR}}\else${v_{\rm LSR}}$\fi}
\newcommand\Tmb{{T_{\rm MB}}}
\newcommand\Ta {{T_{\rm A}^*}}
\newcommand\dv {{\rm d}v}
\newcommand\Eu {{E_{\rm u}}}
\newcommand\vexp{{v_{\rm exp}}}
\newcommand\vcol{{v_{\rm col}}}
\newcommand\CO{{\rm CO}}
\newcommand\CN{{\rm CN}}
\newcommand\HOCp{{\rm HOC^+}}
\newcommand\SiO[2]{{\rm {^{#1}Si{^{#2}O}}}}
\newcommand\HHHp{{\rm H_3^+}}

\newcommand\HCN{{\rm HCN}}
\newcommand\HCOp{\ifmmode{\rm HCO^+}\else{$\mathrm{HCO^+}$}\fi}
\newcommand\HCNt{\ifmmode{\rm H{^{13}C}N}\else{$\mathrm{H{^{13}C}N}$}\fi}
\newcommand\HCOpt{{\rm H{^{13}C}O^+}}

\newcommand\Cn{{\rm C^0}}
\newcommand\Cp{{\rm C^+}}
\newcommand\COt{\ifmmode{\rm {^{13}CO}}\else{$\mathrm{^{13}CO}$}\fi}
\newcommand\Ct{\ifmmode{\rm {^{13}C}}\else{$\mathrm{^{13}C}$}\fi}
\newcommand\ammonia{{\rm NH_3}}

\newcommand\NNHp{{\rm N_2H^+}}
\newcommand\JJ[2]{\ifmmode{\mbox{{\it J} = #1\mbox{--}#2}}\else{{\it J} = #1--#2}\fi}
\newcommand\JN[4]{\mbox{{\it $J_N$} = $#1_{#2}$\mbox{--}$#3_{#4}$}}
\newcommand\CIa{\ifmmode{^3}P_1\mbox{--}{^3}P_0\else${^3}P_1\mbox{--}{^3}P_0$\fi}
\newcommand\CIb{{^3}P_2\mbox{--}{^3}P_1}
\newcommand\thecomplex{{\it l} = 1^\circ.3}
\newcommand\FeI{{\rm FeI 6.4 keV}}
\newcommand\Kalp{{\rm K}_\alpha}
\newcommand\RR[4]{R_{#1\mbox{--}#2/#3\mbox{--}#4}}
\newcommand\gl{l}
\newcommand\gb{b}
\newcommand\ICI{I_{\rm [CI]}}
\newcommand\ICOt{I_{\rm ^{13}CO}}
\newcommand\NCn{\Column{\Cn}}
\newcommand\NCOt{\Column{\COt}}
\newcommand\NCO{\Column{\rm CO}}
\newcommand\ICII{I_{\rm [CII]}}
\newcommand\CIIa{{^2}P_{3/2}\mbox{--}{^2}P_{1/2}}
\newcommand\glgb[2]{\left(\gl, \gb\right) = \left({#1}, {#2}\right)} 
\newcommand\glgbp[2]{\gl, \gb\ = {#1}, {#2}} 
\newcommand\dg[2]{{#1}^\circ.{#2}}
\newcommand\glgbd[4]{\left(\gl, \gb\right) = \left(\dg{#1}{#2}, \dg{#3}{#4}\right)} 
\newcommand\sgras{$\mathrm{Sgr A^{*}}$}
\newcommand\CLa{CO$-0.40$$-0.22$}
\newcommand\CLb{CO$-0.30$$-0.07$}
\newcommand\Dv{\sigma_v}

%%%%%%%%%%%%%%%%%%%%%%%%%%%%%%%%%%%%%%%%%%%%%
%%%%%%%%%%%%%%%%%%%%%%%%%%%%%%%%%%%%%%%%%%%%%
%\NeedsTeXFormat{LaTeX2e}[1995/12/01]%
%\ProvidesFile{table1.tex}%
% [2003/12/12 5.2/AAS markup document class]%

%%%%%%%%%%%%%%%%%%%%%%%%%%%%%%%%%%%%%%%%%%%%%

%%%%%%%%%%%%%%%%%%%%%%%%%%%%%%%%%%%%%%
%%\baselineskip=7mm %%!!!!!!!!!!!!!!!!!
%%%%%%%%%%%%%%%%%%%%%%%%%%%%%%%%%%%%%%

\title{High Velocity Compact Clouds in the Sagittarius C Region}
\author{Kunihiko Tanaka}
\email{ktanaka@phys.keio.ac.jp}
\author{Tomoharu Oka}
\author{Shinji Matsumura}
\affil{Department of Physics, Faculty of Science and Technology, Keio University, 3-14-1 Hiyoshi, Yokohama, Kanagawa 223--8522 Japan}

\author{Makoto Nagai}
\affil{Division of Physics, Faculty of Pure and Applied Sciences, University of Tsukuba, 1-1-1 Ten-noudai, Tsukuba, Ibaraki 305--8571 Japan}

\and 

\author{Kazuhisa Kamegai}
\affil{Department of Industrial Administration, Faculty of Science and Technology, Tokyo University of Science, 2641 Yamazaki, Noda, Chiba 278--8510 Japan}

%\keywords{interstellar: matter --- \ion{H}{2} region: \object{Galactic Center, M-0.02-0.07}}

%somechange
 
\begin{abstract}
We report the detection of extremely broad emission toward two molecular clumps in the Galactic central molecular zone.
We have mapped the Sagittarius C complex ($-0^\circ.61 < l < -0^\circ.27$, $-0^\circ.29 < b < 0^\circ.04$) in the HCN {\it J} = 4--3, $\mathrm{^{13}CO}$ {\it J} = 3--2, and $\mathrm{H^{13}CN}$ {\it J} = 1--0 lines with the ASTE 10 m and NRO 45 m telescopes, detecting bright emission with $80\mbox{--}120$\ $\mathrm{km\,s^{-1}}$ velocity width (in full-width at zero intensity) toward  CO$-0.30$$-0.07$ and CO$-0.40$$-0.22$, which are high velocity compact clouds (HVCCs) identified with our previous CO $J$ = 3--2 survey.
Our data reveal an interesting internal structure of CO$-0.30$$-0.07$ comprising a pair of high velocity lobes.
The spatial-velocity structure of CO$-0.40$$-0.22$ can be also understood as multiple velocity component, or a velocity gradient across the cloud.
They are both located on the rims of two molecular shells of about 10 pc in radius.
Kinetic energies of CO$-0.30$$-0.07$ and CO$-0.40$$-0.22$ are $\left(0.8\mbox{--}2\right)\times10^{49}$ erg and $\left(1\mbox{--}4\right)\times10^{49}$ erg, respectively.
We propose several interpretations of their broad emission: collision between clouds associated with the shells, bipolar outflow, expansion driven by supernovae (SNe), and rotation around a dark massive object.
There scenarios cannot be discriminated because of the insufficient angular resolution of our data, though the absence of visible energy sources associated with the HVCCs seems to favor the cloud--cloud collision scenario.
Kinetic energies of the two molecular shells are $1\times10^{51}$ erg and $0.7\times10^{51}$ erg, which can be furnished by multiple SN or hypernova explosions in $2\times10^5$ yr.
These shells are candidates of molecular superbubbles created after past active star formation.
\end{abstract}

%%%%%%%%%%%%%%%%%%%%%%%%%%%%%%%%%%
%
% INTRODUCTION
%
%%%%%%%%%%%%%%%%%%%%%%%%%%%%%%%%%%

\section{INTRODUCTION}
It has long been recognized that the central molecular zone (CMZ) of our Galaxy is in a highly turbulent state.
Large scale surveys in various molecular and atomic lines \citep{Oka1998,Oka2012,Tsuboi1999,Martin2004,Riquelme2010a} show that the CMZ has anomalous physical and chemical conditions characterized by high density ($\sim10^4\ \pcc$), high gas kinetic temperature ($30$--$100\,\kelvin$), large velocity width ($\gtrsim 40\ \kmps$), and high abundance of SiO and complex organic molecules \citep{Requena-Torres2006}.
The velocity dispersions of the molecular clouds in the CMZ are systematically larger by a factor of $\sim 5$ than those expected from the size--line width relation for the Galactic disk \citep{Oka1998,Oka2001a,Miyazaki2000,Shetty2012a}.
There are also a number of spots with even broader emission than the usual CMZ clouds, whose velocity widths are $\sim100\ \kmps$ \citep{Martin-Pintado1997,Tsuboi1999,Oka1999}.
The CO $\JJ{3}{2}$ survey with the Atacama Submillimeter Telescope Experiment (ASTE) 10 m telescope \citep{Oka2007,Nagai2007,Oka2012} detected a number of spots with high CO $\JJ{3}{2}$/$\JJ{1}{0}$ ratio ($> 1.5$), out of which about 70 were such clouds with very large velocity width.
They have locally enhanced velocity dispersions and typically $\lesssim 3$ pc sizes, and in many cases are visible well in high density tracer lines as millimeter HCN, $\HCOp$, and submillimeter CO lines \citep{Oka1999,Tanaka2007,Oka2012}, indicating that they are shocked molecular features.
They were named high velocity compact clouds (HVCCs) in \cite{Oka2007}.

The velocity width of the largest HVCCs is $\sim 100\ \kmps$ in a full-width at zero intensity \citep[FWZI;][]{Oka1999,Oka2001a,Oka2008}. 
It is an open question what supports their fast internal motion, because the majority of the HVCCs lack apparent energy sources in their vicinity.
\cite{Oka2008} hypothesized that one of the most energetic HVCC CO$0.02$$-0.02$ was created through interaction with a series of supernova (SN) explosions in a hidden massive stellar cluster.
\cite{Tanaka2011} found the enhancement of the submillimeter atomic carbon line in this HVCC was similar to that seen in SN-shocked clouds in the Galactic disk \citep{White1994,Arikawa1999}.
\cite{Tanaka2007} investigated another energetic HVCC CO1.27+0.01 in the $l$ = $\dg{1}{3}$ complex and found that it was a part of a multiple expanding shell, which might evolve into a molecular superbubble.
Their hypothesis suggests past micro-starburst activity in those regions.
Alternatively, cloud--cloud collision at a high velocity can also create broad features. 
Such collisions may take place near the $x_1$-$x_2$ orbit intersections \citep{Huettemeister1998,Riquelme2010}, or in the foot points of magnetic floated loops \citep{Fukui2006,Torii2010,Riquelme2013}.

In this paper we report the detection of strikingly broad and bright HCN $\JJ{4}{3}$ emission toward two HVCCs, \CLa\  and \CLb\ \citep{Oka2007,Oka2012}, near the Sgr C giant molecular cloud (GMC) complex.
They were first identified with the ASTE CO $\JJ{3}{2}$ survey \citep{Oka2007,Oka2012}, but their images suffered from severe contamination by surrounding less dense gas and absorption by the foreground Galactic disk, and hence their basic properties such as masses, sizes and velocity widths were unknown.
We successfully obtained their uncontaminated images in full spatial and velocity extent in our HCN $\JJ{4}{3}$ data, owing to its very high critical density ($\sim10^6\ \pcc$).
We also analyze kinematics of two newly found large molecular shells which are likely to be associated with the HVCCs.
Based on these data we discuss several possible origins for the HVCCs, including past local starburst activity, interaction with compact supernova remnants (SNRs), and interaction with dark massive objects.
We adopt 8.3 kpc for the distance to the Galactic center \citep{Gillessen2009}.

%%%%%%%%%%%%%%%%%%%%%%%%%%%%%%%%%%
%
% Observation
%
%%%%%%%%%%%%%%%%%%%%%%%%%%%%%%%%%%

\section{OBSERVATIONS}
\subsection{ASTE 10 m Observation}
The HCN $\JJ{4}{3}$ (354.505 GHz) observations were performed with the ASTE 10 m telescope in 2010 August. 
We made an on-the-fly (OTF) mapping of a $20'\times20'$ region centered at ($\gl$, $\gb$) = ($-0^\circ.45, -0^\circ.17$), including a major part of the Sgr C complex and the gap region between the Sgr A and Sgr C regions.
For convenience we refer to this region as the ``Sgr C complex'' in this paper, although it also contains clouds unassociated with the Sgr C \ion{H}{2} region.
We also made follow-up observations of the \COt\ \JJ{3}{2} line  (330.588 GHz) in 2011 November and mapped $5'\times5'$ regions centered at the HVCCs \CLb\ and \CLa.
The nomenclature of the HVCCs are basically taken from the molecular cloud catalog of \cite{Oka2012}.
\CLa\ was cataloged as CO$-0.41$$-0.23$ in their list, but we use the former nomenclature in this paper because the latter does not represent the correct center position of the cloud.
% its actual center position in the Galactic coordinate is more close to ($\gl$, $\gb$) = ($-0^\circ.40, -0^\circ.22$).

In the ASTE observations a sideband-separation type receiver CATS345 was used as the receiver frontend.  
The telescope beam size at 350 GHz was $22''$. 
The reference position was taken at ($l$, $b$) = ($1^\circ, -1^\circ$).
As the backend, we used a 1024 ch auto-correlator system with a channel separation of 512 kHz corresponding to a 0.44 $\kmps$ velocity separation at 350 GHz.
Typical system noise temperature during the observations was 200--300 K. 
Antenna pointing accuracy was maintained within $5''$ by CO $\JJ{3}{2}$ observations toward V1427 Aql.
The antenna temperatures were calibrated with the standard chopper-wheel method.
The total on-source integration time was 8 hr.
The correction factor for the sideband rejection ratio and the main-beam efficiency was measured to be 1.8, by using M17SW as the intensity calibrator.
The intensity reproducibility was within 10 \% rms during the 11 observation days.
After the subtraction of spectral baselines, the data were convolved with a Gaussian-tapered Bessel function and resampled onto a $8''.5\times8''.5\times 2$\ \kmps\ grid by using the NOSTAR package developed at the Nobeyama Radio Observatory (NRO).
The effective spatial resolution is $24''$.
The rms noise level of the final maps is 0.13 K in $\Tmb$ for the HCN $\JJ{4}{3}$ map, and 0.21 K for the $\COt\ \JJ{3}{2}$ maps.

\subsection{NRO 45 m Observation}
We also made observations of $\HCNt$\ $\JJ{1}{0}$ (86.340 GHz) by using the BEARS 25 multi-beam receiver system on the NRO 45 m telescope in 2010 January.
The same region covered by the ASTE observations was mapped with OTF scans.
The telescope beam size at 86 GHz was $20''$.
The digital backend was operated with the wide-band mode with a channel separation of 512 kHz corresponding to a 1.78 $\kmps$ velocity separation at 86 GHz.
Antenna temperatures were calibrated by the standard chopper-wheel method.  
The correction factor for the main-beam efficiency (1/$\eta_{\rm MB}$) was 2.5. 
Antenna pointing accuracy was maintained within $3''$ during the observation run by observing the SiO $\JJ{1}{0},\ v = 1,2$ maser lines toward VX Sgr.   
We created $\gl$--$\gb$--$\vlsr$ data cube with a $8''.5\times8''.5\times 2\ \kmps$ grid and a $24''$ spatial resolution with the same procedure used for the ASTE data.
The rms noise level of the final map is 0.08 K in $\Tmb$.

%%%%%%%%%%%%%%%%%%%%%%%%%%%%%%%%%%
%
% results
%
%%%%%%%%%%%%%%%%%%%%%%%%%%%%%%%%%%

\section{RESULTS}\label{RESULTS}

%
% Data Overview
%

\begin{figure*}[!pt]
\begin{center}
{
        \epsscale{0.75}
        \plotone{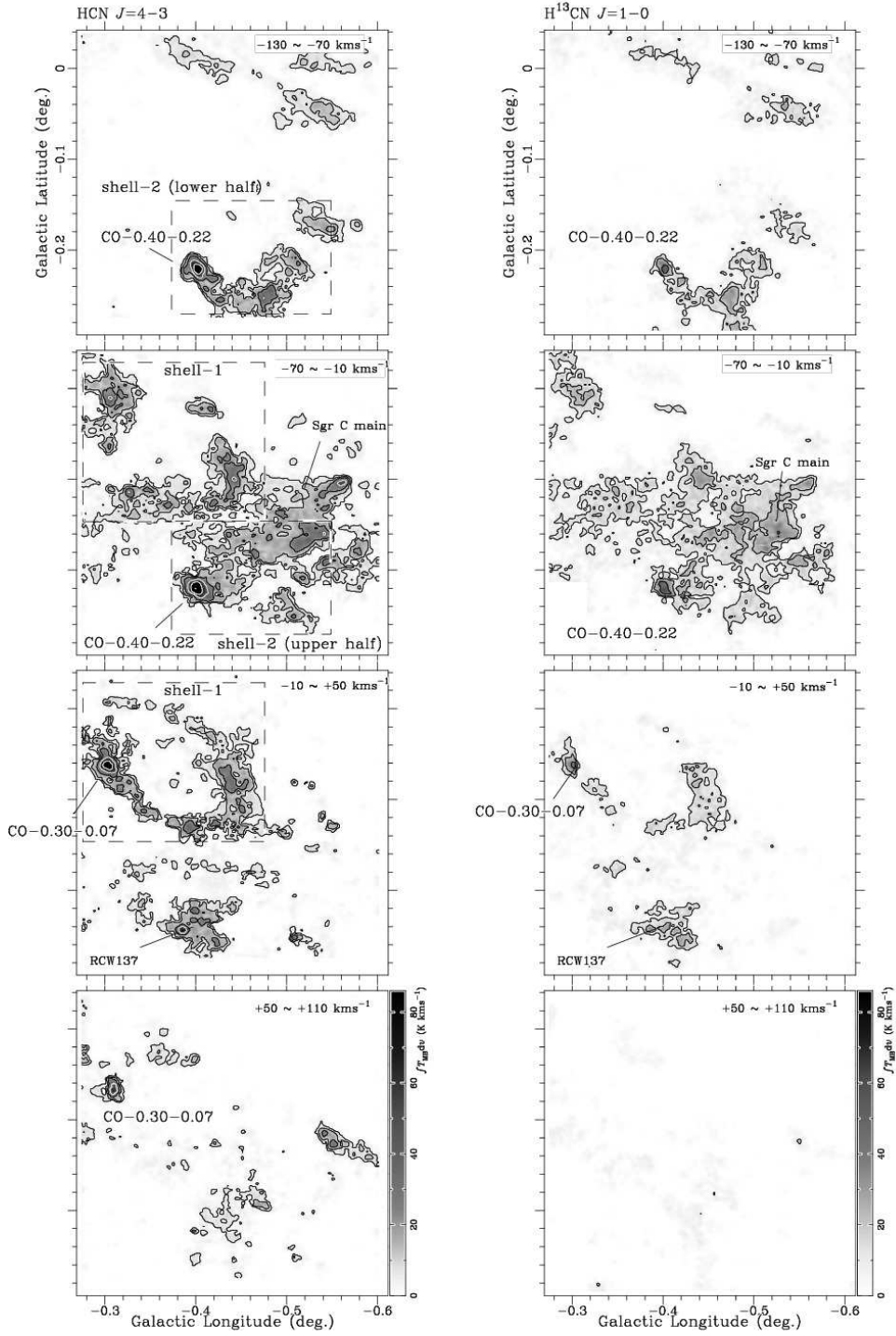}
%        \plotone{f1a.eps}\vspace{5mm}
%        \plotone{f1b.eps}
}
\caption{Velocity channel maps of HCN \JJ{4}{3} (left) and \HCNt\ \JJ{1}{0} (right). Contour levels are 5, 10, 20, 40, 80 and 160 K $\kmps$ for HCN \JJ{4}{3}, and are 10, 20, 40 and 80 K $\kmps$ for \HCNt\ \JJ{1}{0}. }\label{FIG1}
\end{center}
\end{figure*}

Figure \ref{FIG1} shows velocity channel maps of HCN $\JJ{4}{3}$ and $\HCNt$\ $\JJ{1}{0}$ at 60 \kmps\  velocity interval in the range from $-130\ \kmps$ to $+110\ \kmps$. 
This region contains multiple velocity components presumably at different line-of-sight positions \citep{Liszt1995,Lang2010,Riquelme2010a}.
The first and second channels in Figure\,\ref{FIG1} roughly correspond to the $-100\ \kmps$ and $-65\ \kmps$ clouds in the nomenclature in \cite{Lang2010}, or to the clouds 10 and 11 in the nomenclature in \cite{Riquelme2010a}.
The Sgr C \ion{H}{2} region is associated with the latter component \citep{Liszt1995}. 
The cloud in the southern part ($\glgbp{\dg{-0}{4}}{\dg{-0}{2}}$) of the third velocity channel is not a Galactic center source but is associated with the \ion{H}{2} region RCW137 at a distance of 1.5 kpc \citep{Russeil2003}.

\begin{figure*}[!pt]
\begin{center}

{\epsscale{0.75}\plotone{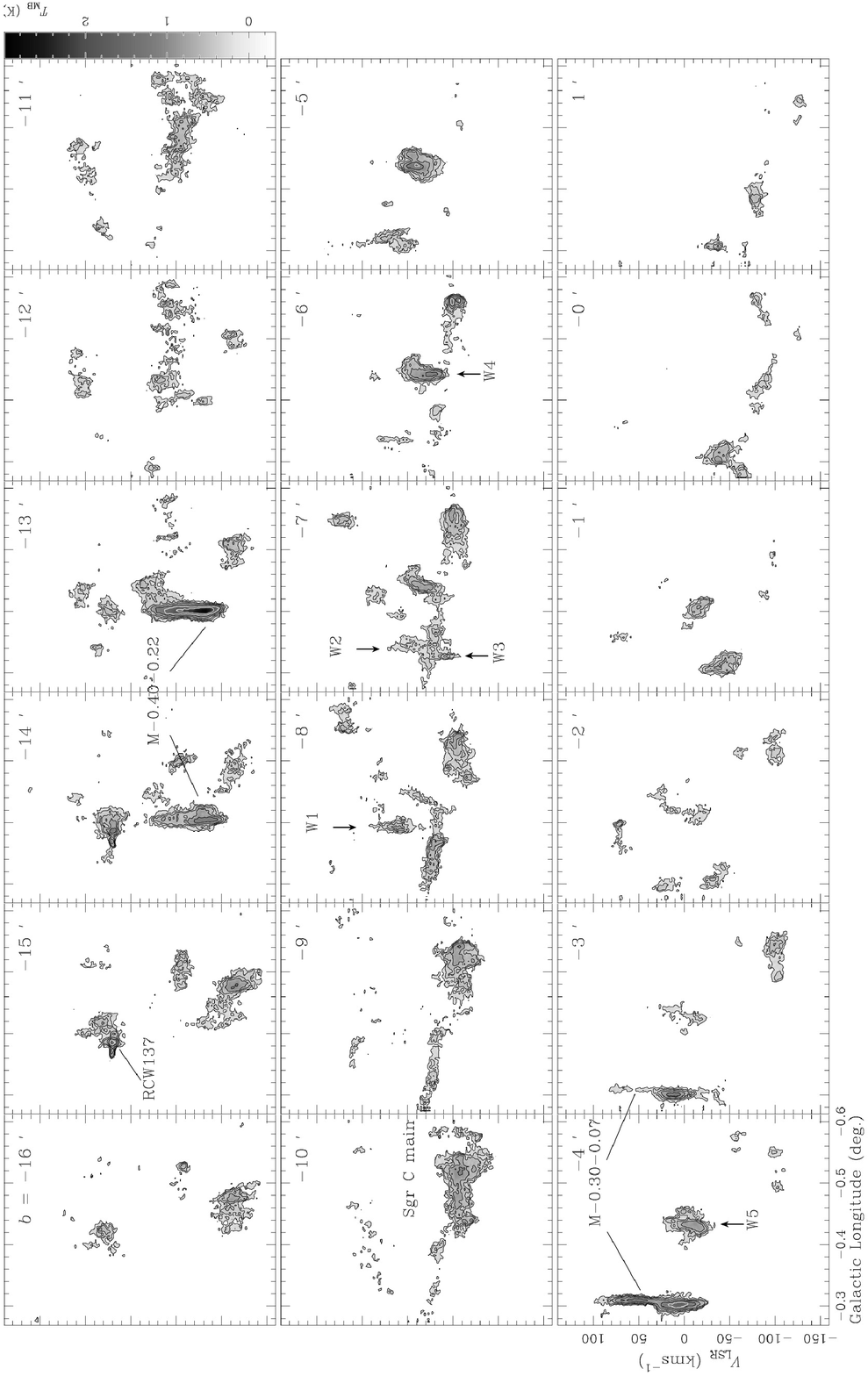}}
\end{center}
\caption{Longitude-velocity diagrams of HCN $\JJ{4}{3}$. Contour levels are 3, 5, 7, 10, 15, 20, and 30 K.  The labels W1--5 denote the positions of the moderately broad line features associated with the shell 1 (Section \ref{RESULTS_SHELLS}).  }\label{FIG2}                
\end{figure*}

\newcommand\okanyoro{CO$0.02$$-0.02$}
\newcommand\tanyoro{CO$1.27$$+0.01$}
We detected remarkably intense emission toward compact clouds \CLa\ and \CLb, whose peak HCN \JJ{4}{3}\ integrated intensities are 530 K \kmps\ and 250 K \kmps, respectively, whereas the intensity does not exceed 120 K \kmps\ even in the densest part of the GMC main component.
They are both identified as HVCCs in \cite{Oka2012} on the basis of their CO \JJ{3}{2}\ survey data, but they are less bright in the CO line and heavily contaminated with less dense, spatially extended gas.
In our HCN \JJ{4}{3}\ map they are well isolated from the main component of the GMC complex and their extraordinarily large velocity width is clearly seen.
Figure \ref{FIG2} shows Galactic longitude-velocity diagrams of HCN $\JJ{4}{3}$ at $1'$ latitude interval.
In the figure \CLa\  ($\gb$ = $-14'$ to $-13'$) and \CLb\  ($\gb$ = $-4'$ to $-3'$) are seen as bright broad line features with $80$\ \kmps\  and $120$\ \kmps\  FWZI, respectively, whereas the velocity width is $\sim20$ \kmps\ for other regions in the map.

Their velocity widths are notably large even compared with other HVCCs identified in \cite{Oka2012}.
Most of the HVCCs detected with CO observations have $\sim 50\ \kmps$ FWZI or smaller velocity width, with only a few exceptions such as \okanyoro\ and \tanyoro\ \citep{Tanaka2007,Oka2008}.
We list the parameters of \CLa\ and \CLb\ in Table 1.

\begin{figure*}[!pt]
\begin{center}
\epsscale{0.85}\plotone{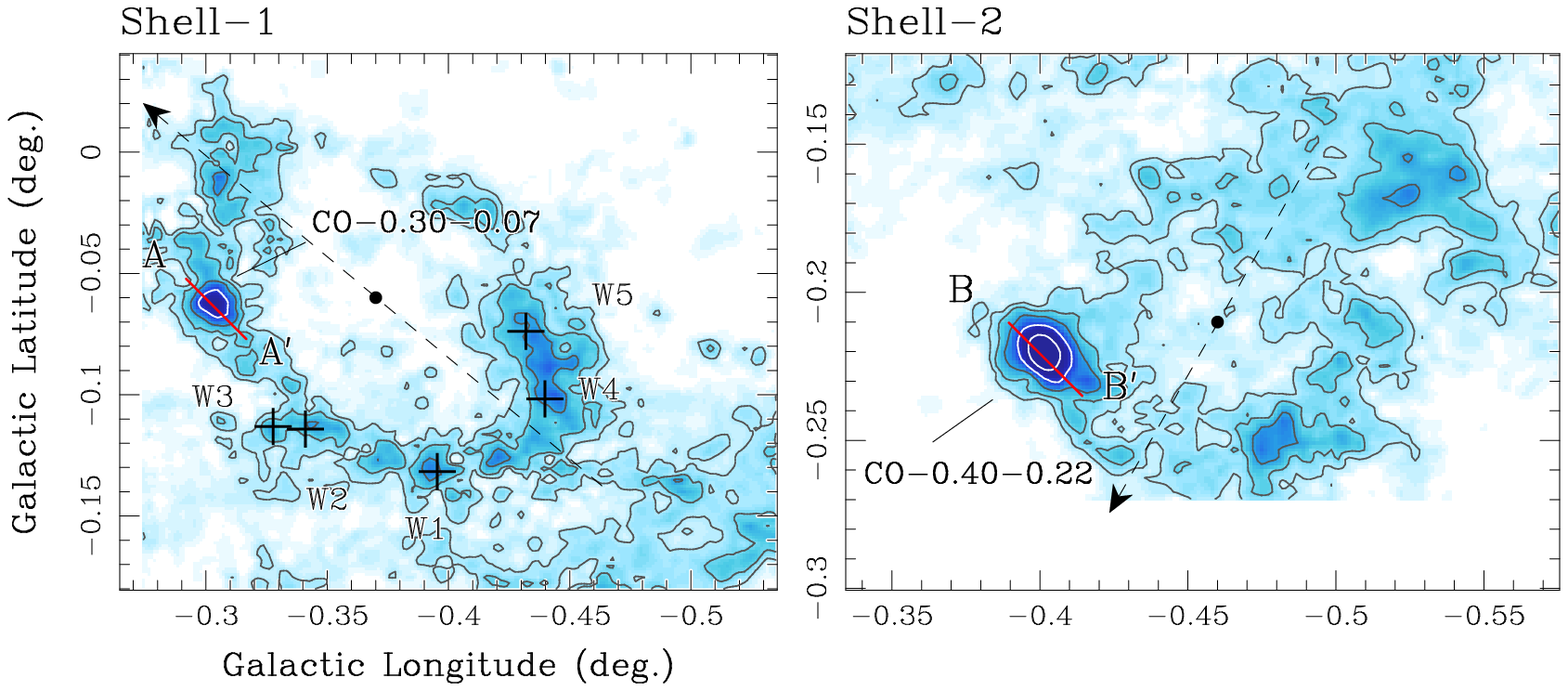}
\caption{HCN \JJ{4}{3}\ integrated intensity maps near (a) the shell 1 and (b) the shell 2.  Integration velocity range is from $-60\ \kmps$ to $50\ \kmps$ for shell 1 and from $-150\ \kmps$ to $-20\ \kmps$ for shell 2.  Contour levels are 5, 10, 20, 40, 80, and 160 K $\kmps$. The labels W1--5 denote the positions of the moderately broad line features (Section \ref{RESULTS_SHELLS}).  The broken lines are the major axis of the shells determined by eye-fitting. The center positions are denoted by dots.  The lines A--A$'$ and B--B$'$ are the strip lines along which the position--velocity diagrams of the HVCCs (Figure\,\ref{FIG4}) are made.}
\label{FIG3}
\end{center}
\end{figure*}

We also find that the HVCCs are located close to two large molecular shells, whose positions are shown in Figure\,\ref{FIG1}.
The shell 1 has a well-defined ellipse morphology with a position angle of $50^\circ$ in the Galactic coordinate.
The HVCC \CLb\ is located on the eastern edge of this shell.
The shell 2 is in the southernmost part of the Sgr C region in the velocity range from $-130$ to $-10\ \kmps$. 
The upper and lower parts of the shell 2 appear in different velocity ranges in Figure\,\ref{FIG1}, and its elliptical shape is more clearly visible in the integrated intensity map shown in Figure\,\ref{FIG3}.
The HVCC \CLa\ is located on the rim of the shell 2.
In Table 1 we listed the center positions, diameters, and position angles of the shells determined with eye-fitting.

\begin{deluxetable}{lcccc}
\input{table1.tex}
\end{deluxetable}

%
% Close-up view of CO-0.30-0.07 and Shell1
%

\begin{figure*}[!pt]
\epsscale{1.}
\plotone{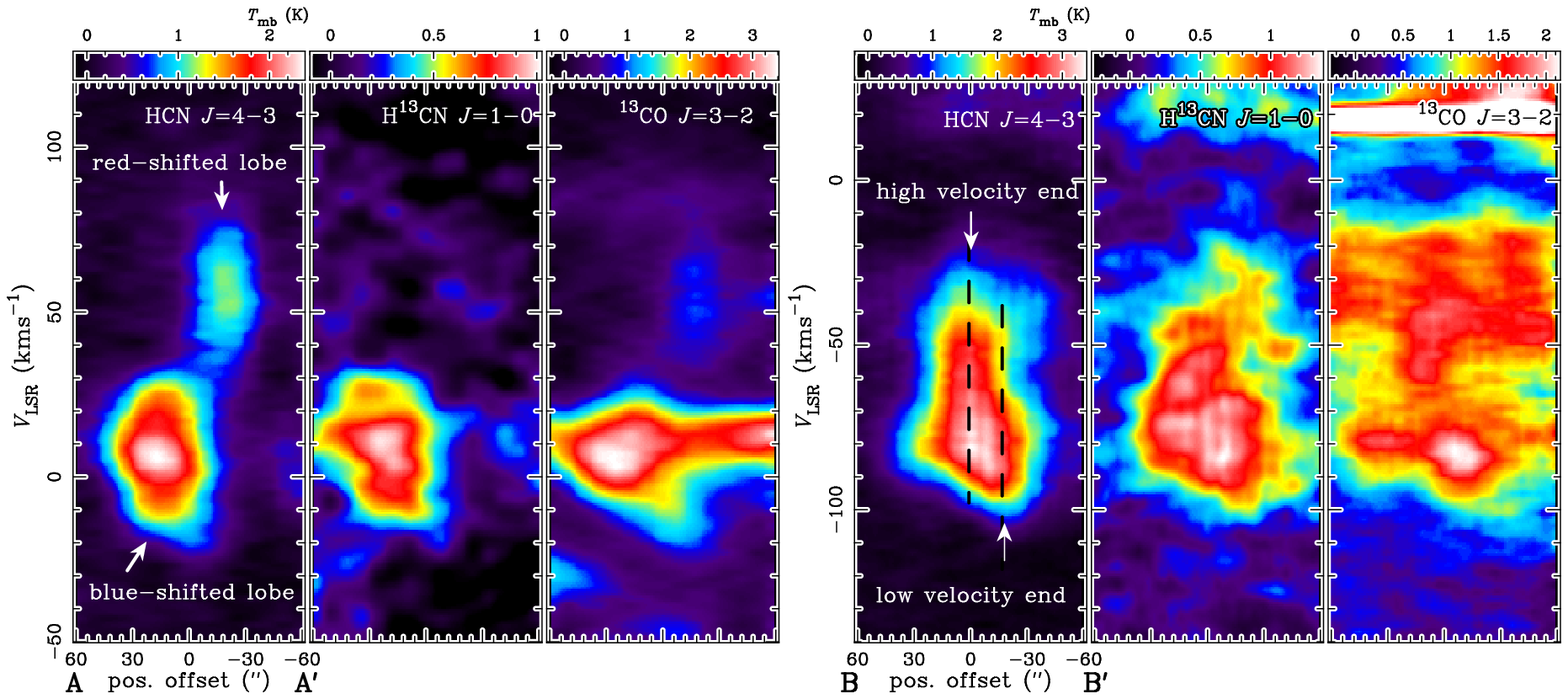}
 \caption{
Position-velocity ($P$-$V$) diagrams of the HVCCs in HCN $\JJ{4}{3}$, \HCNt\ \JJ{1}{0}, and \COt\ $\JJ{3}{2}$, for \CLb\ (left; along the A--$\mathrm{A}'$ cut in Figure\,\ref{FIG3}) and for \CLa\ (right; along the B--$\mathrm{B}'$  cut in Figure\,\ref{FIG3}).
The broad line features composing the HVCCs are the most clearly visible in the HCN $\JJ{4}{3}$ maps, whereas the \COt\ \JJ{3}{2}\ maps are more sensitive to less dense, spatially extended components such as the narrow line feature at $\vlsr = 10\ \kmps$ in the \CLb\ map, and the broad features extended throughout the \CLa\ map.
}\label{FIG4}
\end{figure*}

\subsection{Spatial-Velocity Structure of the HVCCs}
\newcommand\PV{{\it P} -- {\it V}}
Our HCN \JJ{4}{3}\ data reveals an interesting internal spatial-velocity structure of the HVCCs.
Figure \ref{FIG4} presents position--velocity (\PV) diagrams of \CLb\ and \CLa\ in HCN \JJ{4}{3}, \HCNt\ \JJ{1}{0}, and \COt\ \JJ{3}{2} made along the A--$\mathrm{A'}$ and B--$\mathrm{B'}$ cuts shown in Figure\ \ref{FIG3}.
We see that \CLb\ is not a single-profiled clump but a composition of blue- and red-shifted high velocity lobes.
Each lobe has $50\ \kmps$ velocity width in FWZI, and they are sharply separated from each other both spatially and in velocity.
We cannot clearly see the internal structures of the other HVCC, \CLa, but its peak positions at the low and high velocity ends differ by $20''$.
This can be understood as unresolved multiple velocity components or as a velocity gradient across the cloud. 

\begin{figure}[!pt]
\epsscale{0.6}
\plotone{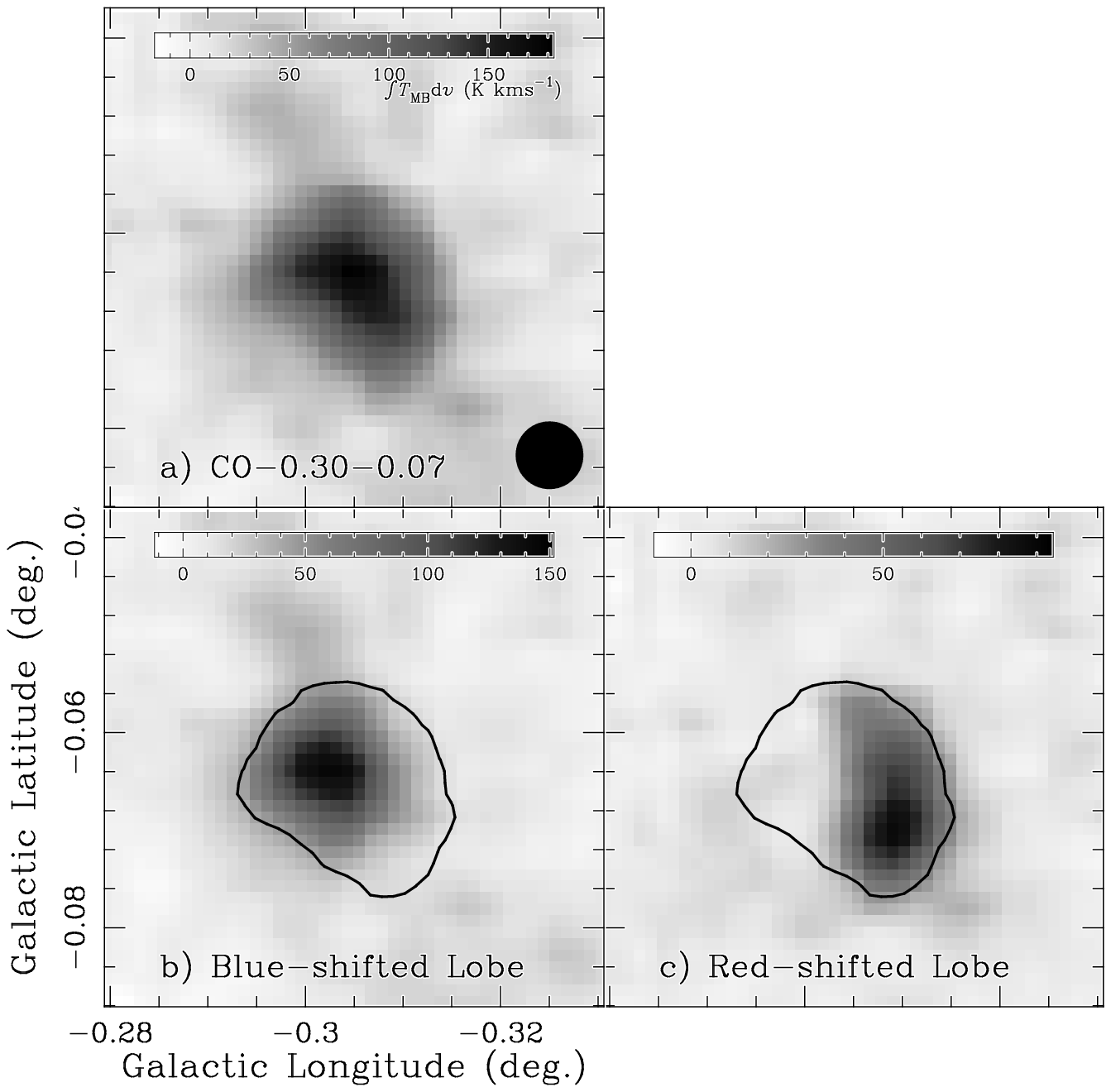}
 \caption{HCN $\JJ{4}{3}$ integrated intensity maps of \CLb, made for the full velocity range (a
; $-30\ \kmps\ \mbox{to}\ +100\ \kmps$), for the blue-shifted lobe (b; $-30\ \kmps$ to $+30\ \kmps$) and for the red-shifted lobe (c; $+40\ \kmps$ to $+100\ \kmps$).  Contours of integrated intensity of $50\ \kelvin\,\kmps$ over the full velocity range are drawn in the panels b and c.  The filled circle on the panel a denotes the effective beam size ($24''$).  
}\label{FIG5}
\end{figure}

It is not likely that the double-lobed \CLb\ is a chance superposition of two unrelated clouds.
Firstly, high density gas is scarce in the $\vlsr$ range higher than 50\ \kmps\ as shown in the Figures \ref{FIG1} and \ref{FIG2}, and therefore it would be unreasonable to assume that a peculiar broad feature in this velocity range falls along the same line-of-sight as another peculiar cloud by chance.
In addition, their morphology implies that they are physically related to each other.
Figure \ref{FIG5} shows HCN $\JJ{4}{3}$ integrated intensity maps of the red- and blue-shifted lobes, in which they show a clear spatial anti-correlation. 
In particular, the arched shape of the red-shifted lobe is in a good accordance with the curvature of the right edge of the blue-shifted lobe.
Therefore, it is more plausible that they are interacting clouds, or represent different portions of a systematic motion such as expansion or rotation. 

The compact broad line features are also visible in \HCNt\ \JJ{1}{0}\ and \COt\ \JJ{3}{2}\ but less clearly, for the both HVCCs.
The \COt\ line has a relatively low critical density of $10^4\ \pcc$, and hence is more sensitive to less dense, quiescent molecular gas invisible in HCN $\JJ{4}{3}$, such as the narrow line component at $\vlsr\sim 10\ \kmps$ near \CLb\ (Figure \ref{FIG4}).
In particular the \COt\ \JJ{3}{2} map of \CLa\ is dominated by spatially extended broad emission of the shell 2, whose velocity extent is equal to that of the HVCC. 
The \HCNt\ map has intermediate characteristics between those of HCN $\JJ{4}{3}$ and \COt: isolated broad feature is present but is spatially more extended than in HCN.

\subsection{Images of the HVCCs in Other Wavelengths}\label{RESULTS2_5}
\begin{figure}[!pt]
\begin{center}
{\epsscale{0.8}\plotone{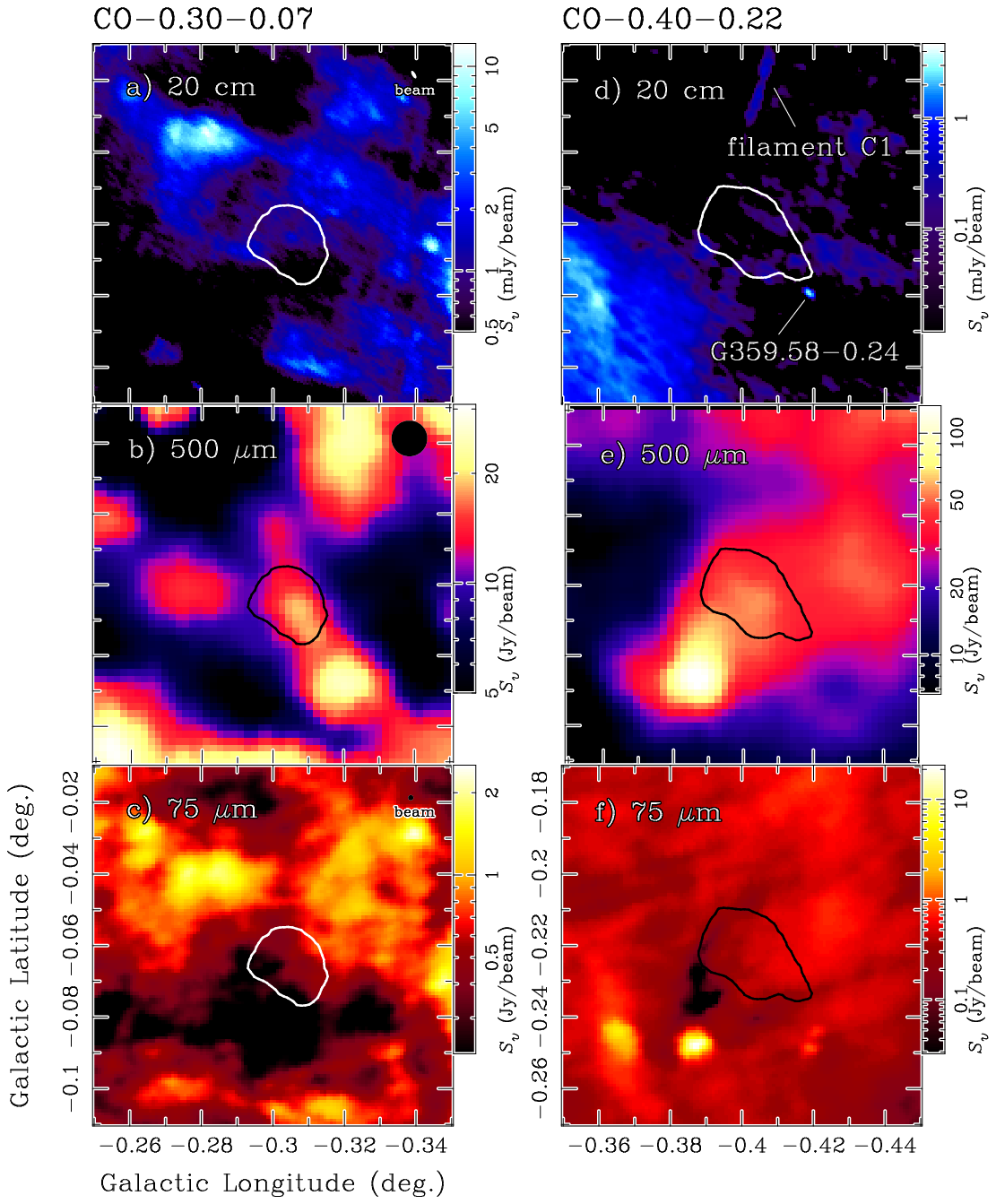}}
\caption{Continuum maps of 20 cm, 500 $\micron$, and 75 $\micron$ wavelengths near \CLb\ (left)  and  \CLa\ (right).
The 20 cm maps are taken from the VLA archival data, and the 500 and 75 $\micron$ maps are from the {\it Herschel} archival data.  The contours are drawn at an HCN $\JJ{4}{3}$ integrated intensity level of 50 K$\kmps$. Spatial resolution is denoted by a filled circle for each map.  We see counterparts of the HVCCs in the 500 \micron\ images, but not for the 75 $\micron$ and the 20 cm images.}\label{FIG6}
\end{center}
\end{figure}

Currently no bright continuum sources are known toward \CLa\ and \CLb.
Figure \ref{FIG6} shows radio and far-infrared continuum ($\lambda = 20\ \mathrm{cm}, 75\ \micron$, and 500 $\micron$) maps near \CLb\ ad \CLa\ taken from the data archives of the Very Large Array (VLA) and the {\it Herschel Space Observatory}.
The 500 \micron\ emission traces the column density of the cold dust \citep[$T\sim20$ K; ][]{Pierce-Price2000}.
We see small 500 \micron\ emission excess both toward \CLb\ and toward \CLa. 
This dimness in 500 \micron\ indicates that the very bright HCN emission of the HVCCs is because of high excitation temperatures rather than because of high column density of the clouds.
The 20 cm and 75 \micron\ maps show neither of \CLb\ nor \CLa\ are associated with bright thermal or non-thermal compact sources indicative of young massive stars or SNRs.
The bright sources in 20 cm and 75 \micron\ to the southeast of \CLa\ is from the foreground \ion{H}{2} region RCW137.
Two 20 cm sources are also seen in the vicinity of \CLa, a non-thermal radio filament C1 and a point source G$359.58$$-0.24$ with unknown nature \citep{YusefZadeh2004}, but no evidence to imply their association to the cloud is found. 

\subsection{Expansion Motion of the Shells}\label{RESULTS_SHELLS}

\begin{figure}[!pt]
\begin{center}
{\epsscale{0.75}\plotone{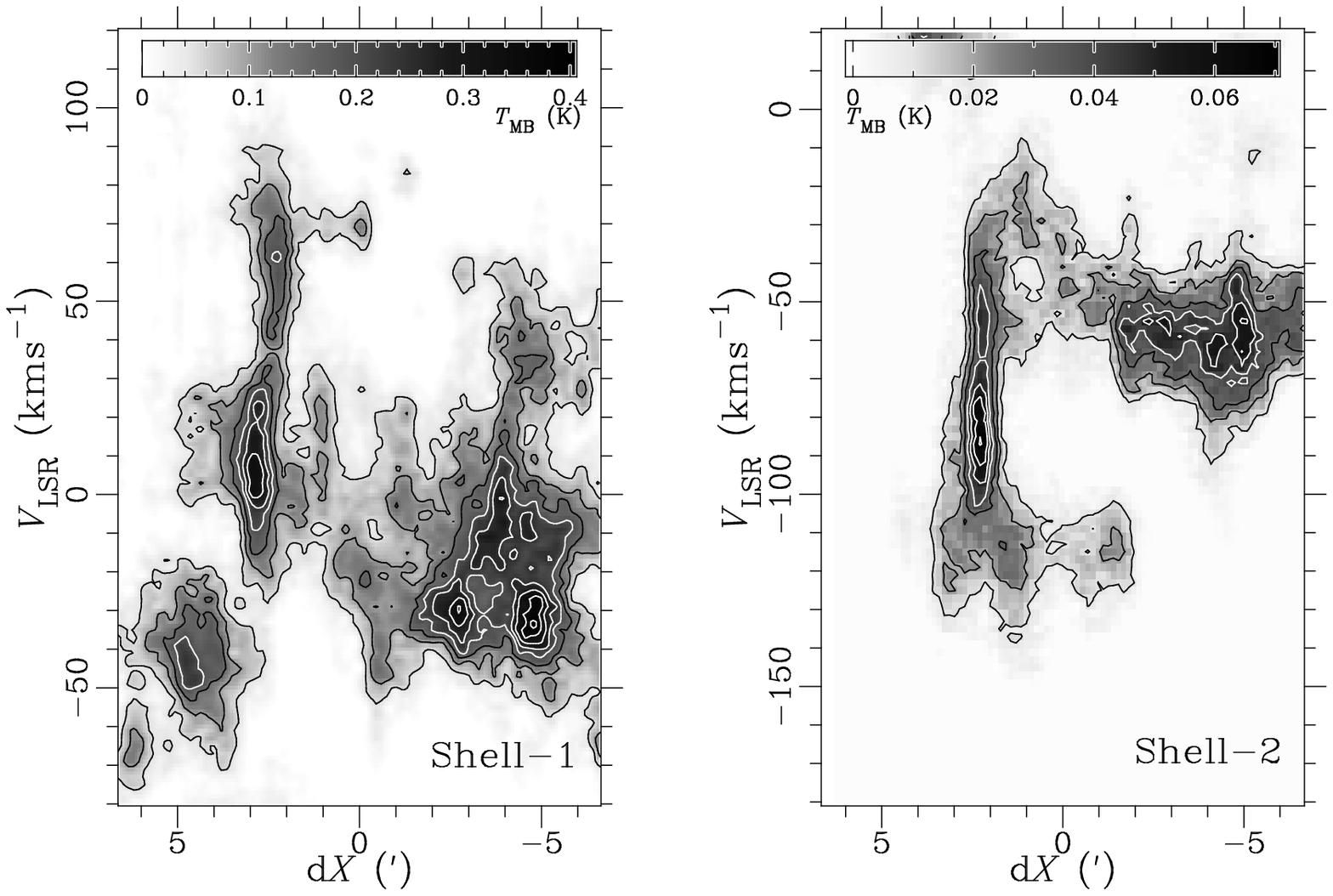}}
\end{center}
\caption{Position--velocity diagrams of the shell 1 (left) and the shell 2 (right) made along their major axes.  Emission is averaged over the full extent in the direction of the minor axes.   The relative distance $\mathrm{d}X$ is measured from the center position of the shell in the direction denoted by the arrows in Figure \ref{FIG3}. Contours are drawn at every 0.05 K for the shell 1, and at every 0.01 K for the shell 2.  }\label{FIG7}
\end{figure}

\newcommand\dt{\theta}
Figure \ref{FIG7} shows \PV\ diagrams of the shells 1 and 2 in HCN $\JJ{4}{3}$, cut along their major axes (shown in Figure\,\ref{FIG3}) and averaged over the minor axis extent.
Both shells have very large velocity extent of about 100 \kmps.
In addition to the HVCCs \CLb\ and \CLa\ at the left edges, a very broad feature is present at the right edge of the shell 1, whose width is comparable to that of the HVCCs.
This large velocity feature shows curvature in the \PV\ space, which could be understood as a part of an elliptical pattern expected for a shell with a radial expansion motion.
The expansion velocity is roughly estimated to be $40\ \kmps$.
However, the left edge of the shell, i.e., \CLb\ does not have a curvature consistent with this expanding shell model, and it is unclear whether the entire shell has an expansion motion.
On the other hand, the entire part of the shell 2 can be relatively well fitted by a partial elliptical pattern, except for the velocity higher than $-40\ \kmps$, where the higher velocity part of \CLa\ does not follow the curvature of the partial ellipse.
The expansion velocity of the shell 2 is estimated to be $35\ \kmps$.

In addition to the above broad velocity features,  moderately broad features of about $40$--$50\ \kmps$ velocity width are associated with the shell 1 (Figure \ref{FIG2}).
These moderately broad emission features are denoted by labels W1--5 in Figures\,\ref{FIG2} and \ref{FIG3}.
The features W3, 4, and 5 form the broad emission at the right edge of the shell 1 in Figure\,\ref{FIG7}.
We note that the moderately broad emission features appear exclusively on the rim of the shell 1.

\subsection{Physical Conditions of HVCCs}
The noteworthy HCN $\JJ{4}{3}$ brightness of the HVCCs indicates their unusual physical conditions.
We calculate the $\Tkin$ and $\nH$ ranges to reproduce the observed HCN $\JJ{4}{3}$/\HCNt\ \JJ{1}{0}\ ratios of \CLb\ and \CLa, which are 2.6 and 2.5, respectively, by employing the large velocity gradient (LVG) model \citep{Goldreich1974}.
We assume that the [HCN]/[$\HCNt$] isotopic abundance ratio is 24 \citep{Langer1990,Langer1993} and calculate the intensity ratio for the $\mathrm{d}N_\HCN/\mathrm{d}v$ range of $10^{13 \mbox{--} 15}$\ $\psc\left(\kmps\right)^{-1}$.
The rate coefficients are taken from the Leiden atomic and molecular database \citep{Schoier2005}.
The calibration error of 10\ \% in the line intensities are considered in the calculations.
We also perform the same calculation for the `quiescent component' for purpose of comparison, which we define as the pixels that do not belong to either of the HVCCs, the two molecular shells, or the foreground RCW137 region.
Calculations are performed separately for the $-65\ \kmps$ and $-100\ \kmps$ clouds of the quiescent component, whose representative HCN \JJ{4}{3}/\HCNt\ \JJ{1}{0} ratios are 0.8 and 0.6, respectively.

\begin{figure}[!ht]
\begin{center}
\plotone{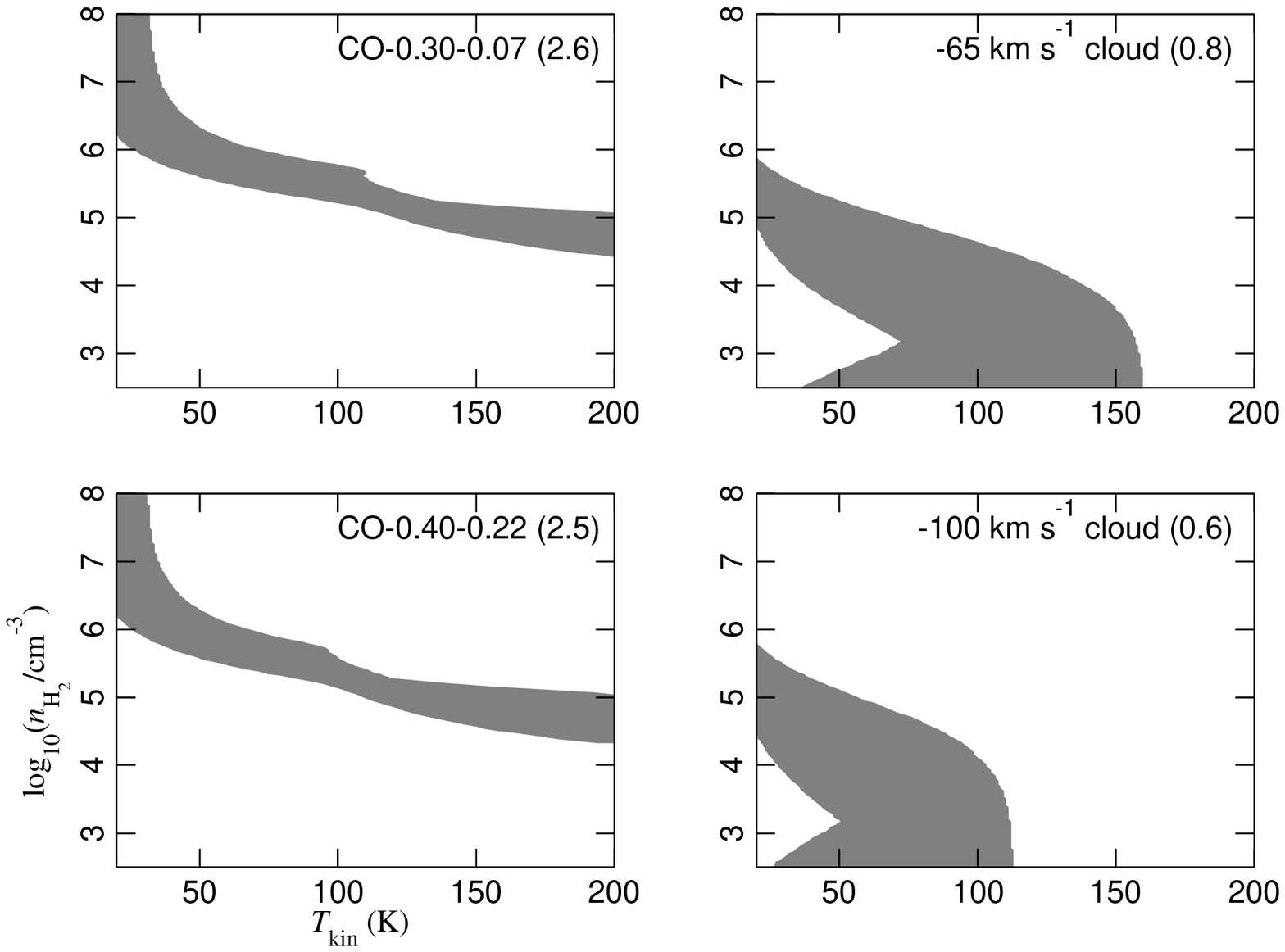} 
\end{center}
\caption{Ranges of $\Tkin$ and $\nH$ calculated with LVG analysis of the HCN \JJ{4}{3}/\HCNt\ \JJ{1}{0} intensity ratios for the HVCCs (\CLb\ and \CLa) and the quiescent clouds ($-65\ \kmps$ cloud and $-100\ \kmps$ cloud).   The shaded areas are the parameter ranges which reproduce the observed HCN \JJ{4}{3}/\HCNt\ \JJ{1}{0}\ ratios (written in the parenthesis) within 10\ \% accuracy.  A $\mathrm{d}N_\HCN/\mathrm{d}v$ range of $10^{13 \mbox{--} 15}$\ $\psc\left(\kmps\right)^{-1}$ is assumed in the calculations. }\label{FIG8}
\end{figure}

The results are shown in Figure\,\ref{FIG8}.
The physical conditions of the quiescent component are consistent with those of the typical CMZ clouds from  preceding studies \citep[e.g.][]{Martin2004,Nagai2007} in that $\nH = 10^{3.5\mbox{--}4}$\ $\pcc$ and $\Tkin = 30$--100 K.
The HVCCs have much higher density and possibly higher temperature than them.
When compared at a $\Tkin$ of $50\ \kelvin$, the density of the $-65\ \kmps$ cloud is less than $10^5\ \pcc$, whereas the density of HVCCs is  $10^{5.5\mbox{--}6.5}\ \pcc$.
Our analysis sets upper limits for the kinetic temperatures of the quiescent component, which are $160\ \kelvin$ and $110\ \kelvin$ for the $-65$ and $-100\ \kmps$ cloud.
On the other hand, much higher temperature of $> 200\ \kelvin$ is allowed for the HVCCs.
This highly enhanced density and possibly temperature, along with the large velocity width, confirms that the HVCCs are shocked features.

\subsection{Mass and Kinetic Energy of the HVCCs and the Shells}\label{RESULTS_MASS}
\newcommand\cfct{X_{\mathrm{HCN}/\mathrm{CO}}}
\newcommand\Td{T_{\mathrm{d}}}
\newcommand\fx{\kelvin\,\kmps\,\mathrm{arcmin}^2}
We estimate masses of the HVCCs from their $\HCN\ \JJ{4}{3}$ and $\HCNt\ \JJ{1}{0}$ luminosity.
If the $\HCNt\ \JJ{1}{0}$ transition is not optically thick, we can obtain a good estimate of the HCN column density from the LVG analysis in the previous section. 
By assuming the [HCN]/[CO] and [\COt]/[$\mathrm{H_2}$] abundance ratios to be $10^{-2.7}$ and $10^{-6}$, respectively \citep{Lis1989,Lis1990,Tanaka2009}, the conversion factor from the $\HCNt\ \JJ{1}{0}$ flux to mass is calculated to be $87\pm3\ \Msol\left(\fx\right)^{-1}$ from their HCN $\JJ{4}{3}$/$\HCNt\ \JJ{1}{0}$ ratios, on the condition that $30\ \kelvin < \Tkin < 200\ \kelvin$, $10^{2.5}\ \pcc < \nH < 10^8\ \pcc$, and the $\HCNt\ \JJ{1}{0}$ optical depth is less than 0.5.
The $\HCNt\ \JJ{1}{0}$ flux is 20 $\fx$ for \CLb\ and 40 $\fx$ for \CLa, which are converted into masses of $2\times10^3\ \Msol$ and $4\times10^3\ \Msol$, respectively.
These masses are similar to those of normal CMZ clouds identified with the CS $\JJ{1}{0}$ survey of \cite{Tsuboi1999}.
The masses of the CS clouds of 3--4 pc in diameter (i.e., similar sizes to \CLa\ and \CLb) are in a range from $5\times10^3\ \Msol$ to $4\times10^4\ \Msol$ \citep{Miyazaki2000}.
The CS luminosity mass of \CLa, or the cloud 32 in their nomenclature, is $5.9\times10^3\ \Msol$, being consistent with our estimate.

We also check the above mass estimates by comparing them with masses estimated from 500 \micron\ luminosity.
It is known that the cold dust temperature is not coupled with the gas temperature in the CMZ, and is relatively low and uniform throughout the region \citep[][and references therein]{Pierce-Price2000}.
By assuming  a dust temperature of 20 K and the dust emissivity index ($\beta$) of 2, 
the mass ($M$) is derived as $M/\Msol = 83\ S_{500}/\mathrm{Jy}$ from 500 $\micron$ flux density ($S_{500}$), including the correction factor of 2 for the high metallicity in the CMZ \citep[][and references therein]{Pierce-Price2000}.
The 500 \micron\ flux density of \CLb\  is measured to be 12 Jy, which gives a mass of $1\times10^3\ \Msol$.
This is a good agreement with the HCN luminosity mass of $2\times10^3\ \Msol$, considering large uncertainties in the dust-to-gas ratio, the [$\HCNt$]/[$\mathrm{H_2}$] abundance ratio, and the dust temperature.

We estimate the kinetic energies of the HVCCs on the basis of the above mass estimates with the equation $\Ekin = \frac{1}{2}\alpha M\Dv^2$, where $\sigma_v$ is mass-weighted velocity dispersion.
The factor $\alpha$ ($>1$) is a correction for the projection effect.
The values of $\Dv$ of \CLb\  and \CLa\  are 20 \kmps\ and 18 \kmps\ in \HCNt\ \JJ{1}{0}, respectively, which yields the kinetic energies of $\alpha\cdot0.8\times10^{49}$ erg and $\alpha\cdot1.2\times10^{49}$ erg.
The factor $\alpha$ is 3 for the case of isotropic velocity dispersion, and 1 for the case that the velocity dispersion exists only in the line-of-sight direction.
Hence we estimate the range of the kinetic energy to be $\left(0.8\mbox{--}2\right)\times10^{49}$ erg for \CLb, and $\left(1\mbox{--}4\right)\times10^{49}$ erg for \CLa.

We also estimate the mass of the shells 1 and 2 with the same procedure used for the HVCCs.
The \HCNt\ flux to mass conversion factor for the shells is calculated to be $88\pm2\ \Msol\left(\fx\right)^{-1}$ from the HCN $\JJ{4}{3}$/\HCNt\ \JJ{1}{0}\ ratio of 0.8.
The \HCNt\ flux associated with the shell is $1\times10^3\ \kelvin\,\kmps\,\mathrm{arcmin}^2$ for the shell 1 and $7\times10^2\ \kelvin\,\kmps\,\mathrm{arcmin}^2$ for the shell 2, which are converted into masses of $9\times10^4\ \Msol$ and $6\times10^4\ \Msol$, respectively.
Their kinetic energy of the expansion motion of the shells is given by $\frac{1}{2}M\,\vexp^2$, which is $1\times10^{51}\ \ergs$ for the shell 1 and $0.7\times10^{51}\ \ergs$ for the shell 2 when the $\vexp$ values of $40\ \kmps$ and $35\ \kmps$ are adopted for the shells 1 and 2, respectively.
We list the masses and kinetic energies of the shells in Table 1, along with the results for \CLa\ and \CLb.

\section{DISCUSSION}\label{DISCUSSION}
\subsection{Possible Origins of \CLa\ and \CLb}
We have shown that the HVCCs in the Sgr C complex, \CLa\ and \CLb, have bright HCN $\JJ{4}{3}$ emission and large velocity width of 80--100$\ \kmps$, in spite of their small diameters of about 4 pc.
They lie a factor of 20 above the size--line width relation for the Galactic disk, and about a factor of 4 above that for the CMZ \citep[][and references therein]{Shetty2012a}.
Their velocity widths are notably large even compared with the other HVCCs found with previous CO surveys.
Our HCN $\JJ{4}{3}$ observations have also revealed their internal spatial-velocity structure: the double high velocity lobes of \CLb\ and velocity gradient of \CLa.
We examine the origin of \CLa\ and \CLb, in terms of cloud--cloud collision, bipolar outflow, expansion motion, and rotation.

\subsubsection{Cloud--Cloud Collision}\label{DISCUSSION3}
The double-lobed structure of \CLb\ is readily explained if we assume that it is a pair of colliding clouds.
The velocity gradient of \CLa\ could be also understood as unresolved multiple velocity components.
An advantage of this cloud--cloud collision interpretation is that it does not require energy sources at the positions of the HVCCs, and hence can explain the absence of bright continuum emission associated with them.

The HVCCs are both located on the rim of large molecular shells which have signs of  expansion motion.  
The $\COt\ \JJ{3}{2}$ map of \CLa\ (Figure\,\ref{FIG4}) especially gives the impression that the broad HCN of the HVCCs is a part of a spatially extended \COt\ emission of the shell 2.
We also note that the simple expansion models for the shells deviate from the observations at the positions of the HVCCs.
These situations may imply that the HVCCs are points where the expanding shells are colliding with dense gas in the outside quiescent component.
The maximum turbulent velocity which can be caused by collision is roughly equal to the velocity difference between the colliding clouds, which is expected to be similar to the expansion velocity $\vexp$ of the shell.
The $\vexp$ values of the shells (40\ \kmps\ and 35\ \kmps\ for the shells 1 and 2, respectively) are consistent with the velocity dispersion of the HVCCs, which is $20\ \kmps$ for \CLb\ and $18\ \kmps$ for \CLa.
The moderately broad line features associated with the shell 1 can be also considered as similar collision points associated with the shell.
This scenario is related to another problem of energy sources of the shells 1 and 2, about which we discuss in the Section \ref{DISCUSSION_SHELLS}.

We could consider other processes that cause cloud--cloud collisions associated with large-scale dynamics in the CMZ.
Broad molecular emission is reported at the foot-points of the magnetic floated loops in outer region of the CMZ \citep{Fukui2006,Torii2010}.
However, evidently neither of the HVCCs in the Sgr C complex are associated with loop-like molecular structures.
\cite{Riquelme2010} identified a high velocity cloud at $l\sim1^\circ.3$ to be a cloud falling from the $x_1$ orbit, on the basis of its high $^{12}\mathrm{C}$/$^{13}\mathrm{C}$ isotopic ratio characteristic for the $x_1$ orbit clouds.
We evaluated the $^{12}\mathrm{C}$/$^{13}\mathrm{C}$ isotopic ratio of \CLa, but did not obtain positive results for low $^{13}\mathrm{C}$ abundance.
The $\HCOp\ \JJ{1}{0}$/$\HCOpt\ \JJ{1}{0}$ and the HNC $\JJ{1}{0}$/$\mathrm{HN^{13}C}\ \JJ{1}{0}$ intensity ratios of \CLa\ measured from the data by \cite{Jones2012} are $18\pm3$ and $14\pm3$, respectively, for the $-100\ \kmps$ to $-60\ \kmps$ velocity range of \CLa.
These ratios are lower than the $^{12}\mathrm{C}$/$^{13}\mathrm{C}$ ratio (40--70) measured for the $x_1$ orbit cloud \citep{Riquelme2010} and is consistent with the typical value for the $x_2$ orbit clouds \citep[24;][]{Langer1990,Langer1993}.
For \CLb\ we could not evaluate the isotopic ratios because of an insufficient signal-to-noise ratio.

\subsubsection{Bipolar Outflow}
The appearance of the HVCCs is similar to that of extremely high velocity (EHV) CO emission of molecular outflow sources \citep{Bachiller1990, Leurini2006, Qiu2011}.
There are spots of intense 44.07 GHz methanol emission toward the both HVCCs \citep{Jones2013}, and they are possibly class-I maser sources which are commonly detected toward molecular outflow sources. 
In addition, intense SiO lines, which are well-established tracers of molecular outflows, are also detected toward \CLa\ \citep{Jones2011,Jones2013}.

However, the kinetic energy of the known EHV sources is at most $9\times10^{47}\ \ergs$ \citep{Leurini2006}, which is about an order of magnitude smaller than that of the HVCCs. 
Any YSO candidates are so far not detected in IR images \citep{Yusef-Zadeh2009,Immer2012} nor in OH or class-I methanol maser data \citep{Yusef-Zadeh1999,Caswell2010}.
Therefore, it is unlikely that the HVCCs are bipolar outflow sources, unless we assume a peculiar driving source which is deeply embedded and extremely energetic.

\subsubsection{Compact Expanding Shell}
The double-lobe of the \CLb\ can be also interpreted as an unresolved, compact expanding shell.
In this interpretation, the red- and blue-shifted lobes are assumed to represent the hemispheres with the receding and approaching line-of-sight velocity.
The kinetic energy of the HVCCs of $10^{49}$ erg order is consistent with the energy injected to molecular clouds by a single SN explosion \citep[e.g. ][]{Moriguchi2001,Sashida2013}.
High velocity molecular emission with 80--120 $\kmps$ velocity width is also detected toward IC443 and W44 SNRs \citep{Dickman1992,Sashida2013}.

An argument against this SNR hypothesis is that the HVCCs lack radio and X-ray emission indicative of SNRs.
The dynamical age of the assumed compact shell is $D/\Delta v =  10^4\ \yr$, where $D$ and $\Delta v$ are diameter and FWZI velocity width of the HVCC.
It is an open question whether it is possible that a SNR with this relatively young age does not show detectable emission.
The lack of radio continuum emission may be partly explained by deficiency of relativistic  electrons owing to the strong magnetic field \citep{Oka2001a} and by intense background non-thermal radio emission widespread in the CMZ \citep{LaRosa2005}.

\subsubsection{Rotating Ring/Disk}
We also consider the possibility that the large velocity width of the HVCCs comes from rotation motion.
The double-lobed structure of \CLb\ and the velocity gradient of \CLa\ could be interpreted as the rotation of a ring or disk viewed from edge-on.
The mass enclosed inside a rotating ring with a radius $r$ and a rotation velocity $v_{\mathrm{rot}}$ is $rv_{\mathrm{rot}}^2/G$, where $G$ is the gravitational constant.
When we assume the velocity width of the HVCCs is due to the rotation plus the intrinsic velocity width of $20\ \kmps$, and that the radii are the half distance between the highest and lowest velocity peaks (0.6 pc and 0.4 pc for \CLb\ and \CLa, respectively), the enclosed mass is estimated to be $1\times10^5\ \Msol$ for \CLa\ and $3\times10^4\ \Msol$ for \CLa.
A candidate of a compact massive object with this huge mass is an intermediate-mass black hole (IMBH), which might have been formed through runaway stellar mergers in a compact cluster, or brought from a satellite galaxy by a minor-merger event \citep{Ebisuzaki2001,PortegiesZwart2004}.
Candidates of IMBHs are found as ultra luminous X-ray sources in external galaxies \citep{Colbert1999}, but no bright X-ray sources are known near the HVCCs.

%\vspace{\Cvs}
\subsubsection*{}
The above possible explanations for the large velocity width of \CLa\ and \CLb, namely, collision, bipolar outflow, expansion, and rotation cannot be strictly discriminated with our data because of the insufficient spatial resolution of our data taken with single-dish telescopes.
At present, the absence of visible energy sources at any wavelength seems to favor the collision scenario, because other scenarios require energy sources inside or in the vicinity of the HVCCs.
In particular, in the bipolar outflow and rotating ring/disk scenarios we assume very exotic objects such as extremely energetic outflow-driving stars and IMBHs as the energy sources. 
A more detailed investigation based on interferometric observations would be required to conclude whether the spatial-velocity structure of the HVCCs demands the presence of such objects.

\subsection{Energy Sources of Molecular Shells: Multiple Supernovae/Hypernovae?}\label{DISCUSSION_SHELLS}
The well-defined elliptical structure of the shells and their signs of expansion motion suggest that each of them was created by a point explosion.
A typical core-collapse SN event supplies baryonic energy of $10^{51}\ \ergs$ to interstellar space.
Considering that 0.01--0.1 of the total SN energy is passed into molecular clouds \citep[e.g.][]{Dickman1992,Sashida2013},  about more than 10 SN explosions or two hypernova (HN) explosions could explain the total kinetic energy of the shells.
Dynamical age is $R/\vexp = 2\times10^5\ \yr$ for both shells, and hence this hypothesis requires an SN/HN rate of $10^{-5}$ to $10^{-4}\ \mathrm{yr}^{-1}$.
This SN/HN rate is not an unrealistic value compared with the SN rate of the Galaxy \citep[$1.9\pm1.1\times10^{-2}\ \yr^{-1}$;][]{Diehl2006}, because if we roughly scale the molecular mass to the SN rate, the SN rate of the Sgr C complex is $\left(2\mbox{--}8\right)\times10^{-5}\ \yr^{-1}$ from the 500\ \micron\ luminosity mass of the region ($2.5\times10^6\ \Msol$) and the total Galactic  $\mathrm{H}_2$ mass \citep[$\sim10^9\ \Msol$;][]{Bloemen1986}.

The SN rate gives an estimate of the star formation rate (SFR) of $\sim10^7\ \yr$ ago.
If we assume that 10 SN explosions took place in their dynamical age, and that stars heavier than 8 $\Msol$ cause SN explosions, the past SFR of the Sgr C region is estimated to be $5\times10^{-3}\ \Msol\,{\yr}^{-1}$ on assumption of a Kroupa initial mass function \citep[IMF;][]{Kroupa2001a} with an upper mass cut-off of $50\ \Msol$. 
Alternatively, if we assume that two HN explosions took place in $2\times10^5\ \yr$, i.e., one HN for each shell in their dynamical time, and that  stars heavier than 40 $\Msol$ cause HN explosions, the SFR is $3\times10^{-2}\ \Msol\,{\yr}^{-1}$.
With these assumptions and by using the 500 $\micron$ luminosity mass, the star formation efficiency (SFE) of $10^7\ \yr$ ago is estimated to be $\left(0.2\mbox{--}1\right)\times10^{-8}\,\yr^{-1}$.
This does not reconcile with the present-day SFE of the CMZ \citep[$1\mbox{--}3.4\times10^{-10}\ \yr^{-1}$;][]{Yusef-Zadeh2009}, but is more consistent with the high SFR values at $10^{5\mbox{--}6}\ \yr$ ago suggested by infrared observations \citep[$0.07\mbox{--}0.14\ \Msol\,\yr^{-1}$ for the entire CMZ; ][]{Yusef-Zadeh2009,Immer2012} which yields a SFE value of $\sim1\times10^{-9}\ \yr^{-1}$.

Hence, the large molecular shells 1 and 2 are candidates of molecular superbubbles created after past active star formation, similar to those found in the $\thecomplex$ and Sgr B1 complexes \citep{Tanaka2007,Tanaka2009}.
The Sgr C complex is in fact rich in young stars of ages $10^{5\mbox{--}6}\ \yr$.
\cite{Yusef-Zadeh2009} identified 21 YSOs of $10^5\ \yr$ age in a mass range from $5.4\ \Msol$ to $9.5\ \Msol$ in the 24\ \micron\ map of this region, from which the SFR of the Sgr C region at $10^5\ \yr$ ago is estimated to be $2\times10^{-2}\ \Msol\,\yr^{-1}$ with the Kroupa IMF.
The above SFR value required for the formation of the shells 1 and 2 is not unreasonably high compared with this value.
As shown in our observations, the present Sgr C complex is not rich in high density molecular gas compared to the Sgr A and Sgr B complexes, and is accordingly inactive in star formation except for the vicinity of the Sgr C \ion{H}{2} region. 
The coincidence of the ages of the shells and the YSOs may  mean that the superbubbles cleared up the molecular gas in this region, terminating the active star formation phase which had lasted until $\sim 10^5\ \yr$ ago.

\section{SUMMARY}
We present HCN $\JJ{4}{3}$ and \HCNt\ \JJ{1}{0}\ images of the Sgr C GMC complex taken with the ASTE 10 m and NRO 45 m telescopes.
Below we summarize the main results.

\begin{enumerate}
\item 
We detected strikingly intense HCN $\JJ{4}{3}$ emission toward the two compact clouds, \CLb\ and \CLa, whose integrated intensities are more than a factor of 2 brighter than that in any other part in the complex.
They are HVCCs with very large velocity widths of $\sim 100\ \kmps$ FWZI.
\CLb\ has a distinctive double-lobed structure with a $120\ \kmps$ velocity width.
\CLa\ has an $80\ \kmps$ velocity width, whose spatial-velocity structure can be also understood as multiple velocity component or velocity gradient.
\item We also found two shells of about 10 pc radius each.
They both consist of very broad molecular emission with signs of expansion motion.
If they have expansion motion, the radial expansion velocity is estimated to be $40\ \kmps$ and $35\ \kmps$ for the shells 1 and 2, respectively.
\item 
We made an LVG analysis with the HCN $\JJ{4}{3}$/$\HCNt\ \JJ{1}{0}$ intensity ratio, finding that the HVCCs have a higher density by an order of magnitude and possibly higher temperature than normal clouds in the CMZ. 
\item
Masses of \CLb\ and \CLa\ are estimated to be $2\times10^3\ \Msol$ and $4\times10^3\ \Msol$, respectively, from the \HCNt\ \JJ{1}{0}\ and HCN \JJ{4}{3}\ fluxes.
Their kinetic energies are estimated to be $\left(0.8\mbox{--}2\right)\times10^{49}\ \erg$ and $\left(1\mbox{--}4\right)\times10^{49}\ \ergs$.
Kinetic energies of the molecular shells 1 and 2 are $1\times10^{51}\ \erg$ and $0.7\times10^{51}\ \erg$, respectively.
\item 
We examined several scenarios of the origin of the HVCCs, including cloud--cloud collisions caused by the expansion of the molecular shells, bipolar outflow from young massive stars, compact expanding shells driven by interaction with young SNRs, and ring/disks rotating around dark massive objects.
The spatial resolution of our data is not sufficiently high to discriminate these scenarios, though absence of visible energy sources associated with the HVCCs may favor the cloud--cloud collision scenario.
\item 
More than 10 SN or two HN explosions in the last $2\times10^5\ \yr$ are required to explain the kinetic energy of the expansion motion of each shell.
This SN rate is interpreted into SFR of $\left(0.5\mbox{--}3\right)\times10^{-2}\ \Msol\ \yr^{-1}$ at $10^7\ \yr$ ago, which is about an order higher than the present SFR, but is close to the value in $10^{5\mbox{--}6}\ \yr$ ago.

\end{enumerate}

The authors are grateful to the NRO staffs for their generous support for our observations with the NRO 45 m and the ASTE 10 m telescopes.
We also thank the anonymous referee, whose comments and suggestions were helpful in improving the paper.

     %       
% references
%

\input{references.tex}

%\bibliographystyle{apj}
%\bibliography{mendeley,local}

%
% figures
%

\end{document}

%% file: table1.tex
%\documentclass[preprint]{aastex}

%\input{defines.tex}
%\usepackage{longtable}
%\usepackage{multirow}
%\begin{document}
%\begin{deluxetable}{lcccc}

%%%%%%%
\tablecolumns{5}
\small
\tablecaption{Parameters of HVCCs and Shells}\label{TAB1}
\tablehead{ & \colhead{\CLb} & \colhead{\CLa} & \colhead{Shell-1} & \colhead{Shell-2}}
%\tablewidth{290pt}
\tablewidth{0pt}
\tabletypesize{\scriptsize}
\startdata
Center ($\gl, \gb$) &  $(-0^\circ.30, -0^\circ.07)$ &  $(-0^\circ.40, -0^\circ.22)$ & $(-0^\circ.37, -0^\circ.06)$ & $(-0^\circ.46, -0^\circ.21)$ \\
Diameter ($\mathrm{pc}$) & 3.6 & 4.3 & $24 \times 16$ & $16 \times 12$\\
Position Angle & --- & --- & $50^\circ$ & $-30^\circ$\\
Velocity Dispersion (\kmps)\tablenotemark{a} & 20 & 18 & --- & --- \\
Expansion Velocity (\kmps)\tablenotemark{b} & --- & --- & (40) & 35 \\
\\
Mass ($\Msol$)\tablenotemark{c} & $2\times10^3$ & $4\times10^3$ & $9\times10^4$ & $6\times10^4$ \\
Kinetic Energy ($\erg$)\tablenotemark{d} & $\left(0.8\mbox{--}2\right)\times10^{49}$ & $\left(1\mbox{--}4\right)\times10^{49}$ & $1\times10^{51}$ & $0.7\times10^{51}$ 
\enddata
\tablenotetext{a}{From \HCNt\ \JJ{1}{0} data.}
\tablenotetext{b}{See Section 3.3.}
\tablenotetext{c}{From \HCNt\ \JJ{1}{0} and HCN \JJ{4}{3}\ luminosity. See Section 3.5.}
\tablenotetext{d}{See Section 3.5.}

%%%%%%%%%%

%\end{deluxetable}
%\end{document}

%% file: references.tex
% \newcommand{\noop}[1]{}